\RequirePackage{lineno}
\documentclass[preprint,12pt,square,comma,number]{elsarticle}
\usepackage{CJKutf8}
\usepackage{color,amsmath,amssymb,graphicx,latexsym,subfigure}
\usepackage{threeparttable,multirow,txfonts,ulem,url,float}

\begin{document}

\title{BGO quenching effect on spectral measurements of cosmic-ray nuclei in DAMPE experiment}

\author[1,2]{Zhan-Fang Chen}
\author[1,2]{ Chuan Yue \corref{cor}}
\ead{yuechuan@pmo.ac.cn}
\author[2]{Wei Jiang}
\author[2]{Ming-Yang Cui}
\author[1,2]{Qiang Yuan}
\author[3]{Ying Wang}
\author[3]{Cong Zhao}
\author[3]{Yi-Feng Wei}

\cortext[cor]{Corresponding author}

\address[1]{School of Astronomy and Space Science, University of Science and Technology of China, Hefei 230026, China}
\address[2]{Key Laboratory of Dark Matter and Space Astronomy, Purple Mountain Observatory, Chinese Academy of Sciences, Nanjing 210023, China}
\address[3]{State Key Laboratory of Particle Detection and Electronics, University of Science and Technology of China, Hefei 230026, China}

\begin{abstract}

The Dark Matter Particle Explorer (DAMPE) is a satellite-borne detector designed to measure high energy cosmic-rays and $\gamma$-rays. As a key sub-detector of DAMPE, the Bismuth Germanium Oxide (BGO) imaging calorimeter is utilized to measure the particle energy with a high resolution. The nonlinear fluorescence response of BGO for large ionization energy deposition, known as the quenching effect, results in an under-estimate of the energy measurement for cosmic-ray nuclei. In this paper, various models are employed to characterize the BGO quenching factors obtained from the experimental data of DAMPE. Applying the proper quenching model in the detector simulation process, we investigate the tuned energy responses for various nuclei and compare the results based on two different simulation softwares, i.e. GEANT4 and FLUKA. The BGO quenching effect results in a decrease of the measured energy by approximately $2.5\%$ ($5.7 \%$) for carbon (iron) at $\sim$10 GeV/n and $<1\%$ above 1 TeV/n, respectively. Accordingly, the correction of the BGO quenching effect leads to an increase of the low-energy flux measurement of cosmic-ray nuclei.
\end{abstract}


\date{\today}

\maketitle

\section{Introduction}

Via precisely measuring the energy spectra, mass compositions, arrival directions, and temporal variations of cosmic rays, researchers aim to uncover the mysteries of the origin and propagation of cosmic rays. In the past two decades, a number of experiments have been carried out in high-altitude balloons, orbiting satellites, and space stations. These direct detection experiments can be roughly classified into two types, the magnetic spectrometer (such as PAMELA \cite{adriani2014pamela} and AMS-02 \cite{aguilar2021alpha}) and calorimeter experiments (such as CALET \cite{marrocchesi2012calet}, NUCLEON \cite{atkin2017first}, DAMPE \cite{chang2017dark}, and ISS-CREAM \cite{seo2014cosmic}).
These experiments directly observe cosmic rays with kinetic energies ranging from MeV/n to tens of TeV/n and found that the energy spectra of many species deviate from simple power-laws \cite{adriani2014pamela,atkin2017first,dampe2017direct,2019SciA....5.3793A,alemanno2021measurement,dampe2022detection,aguilar2021alpha,torii2019calorimetric,yoon2017proton,2009BRASP..73..564P}, triggering many discussions on the physics about cosmic rays (e.g., \cite{2012ApJ...752...68V,2012ApJ...752L..13T,2021FrPhy..1624501Y,2023FrPhy..1844301M}). 

The BGO calorimeter is a crucial sub-detector of DAMPE, providing a high accuracy in measuring the energy depositions of incident particles\cite{2022SciBu..67..679C}. However, it suffers from energy leakage due to its limited thickness, which is around 1.6 nuclear interaction lengths.  To evaluate the instrumental response of incident particles in the DAMPE detector, extensive Monte Carlo simulations were conducted using GEANT4 and FLUKA simulation software\cite{allison2016recent,battistoni2015overview}. FLUKA 2011.2x and GEANT 4.10.05 were utilized in this study, as they are widely recognized and have been extensively validated \cite{2020ChPhL..37k9601J}. FLUKA is a comprehensive particle physics MC simulation package that employs the PEANUT package to simulate reactions across the complete energy range. Additionally, FLUKA utilizes the activated DPMJET-III package to simulate hadronic collisions at high energy and nucleus-nucleus collisions above 5 GeV/n. On the other hand, GEANT4 employs the FTFP-BERT physics model to simulate inelastic hadron-nuclear interactions from low to high energy, up to 100 TeV.

The light yield of BGO is expected to be linearly proportional to the energy loss of incident particles when the line energy density ($dE/dx$) is small. However, for a particle with a large charge number, it can induce a large amount of ionization energy loss over a short distance, corresponding to a large $dE/dx$, which may induce a non-linear relationship between the amount of scintillation photons and the energy deposition in the BGO crystal. This effect is known as quenching \cite{birks1951scintillations,brooks1979development}. The quenching effect can lead to an underestimation of the true energy deposition of the particle shower. Previous studies have suggested the evidences of the nonlinearity in the case of light ions or at lower energy levels \cite{avdeichikov1994light,2015NIMPB.360..129T,2021NIMPA.98864865P,valtonen1990response,dlouhy1992response}. Recently, Wei et al. \cite{2020ITNS...67..939W,2019NIMPA.922..177W} investigated the quenching effect of the BGO calorimeter of DAMPE with various nuclei from helium to iron. By comparing the ionization energy losses from the experimental data of DAMPE with that from the simulations, they obtained the BGO quenching factors in a large $dE/dx$ range. 

In this study, we firstly utilize four different quenching models \cite{birks1951scintillations,wright1953scintillation,voltz1966influence,2015NIMPB.360..129T,2021NIMPA.98864865P} to fit the BGO quenching factors of DAMPE and apply the most appropriate model to the GEANT4 and FLUKA simulations to achieve a more precise energy response. By including the BGO quenching effect in the detector simulation, we investigate the tuned energy responses of various nuclei and compared the outcomes from GEANT4 and FLUKA. Finally, we estimated the impact of the quenching effect on the spectral measurement for different nuclei.

\section{Quenching models}
\begin{figure*}
\centering
\includegraphics[width=0.49\textwidth]
{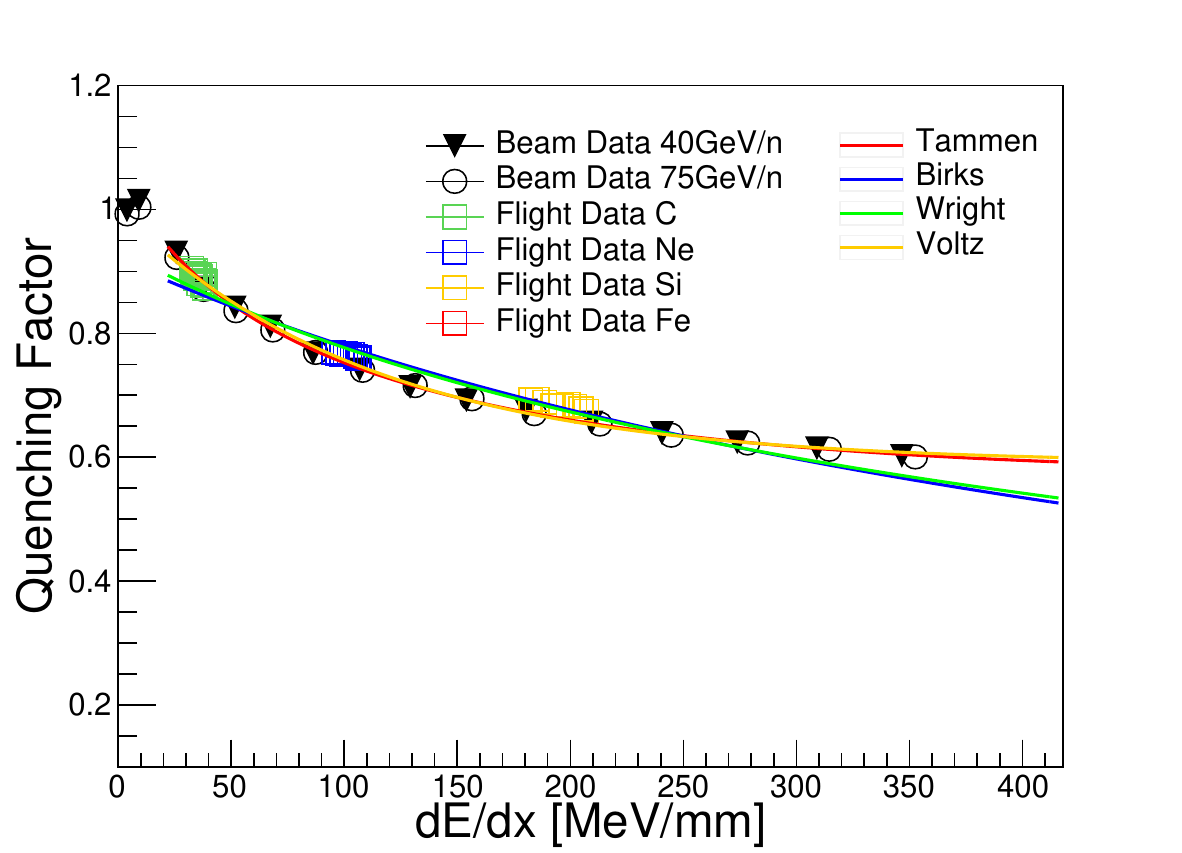}
\includegraphics[width=0.49\textwidth]
{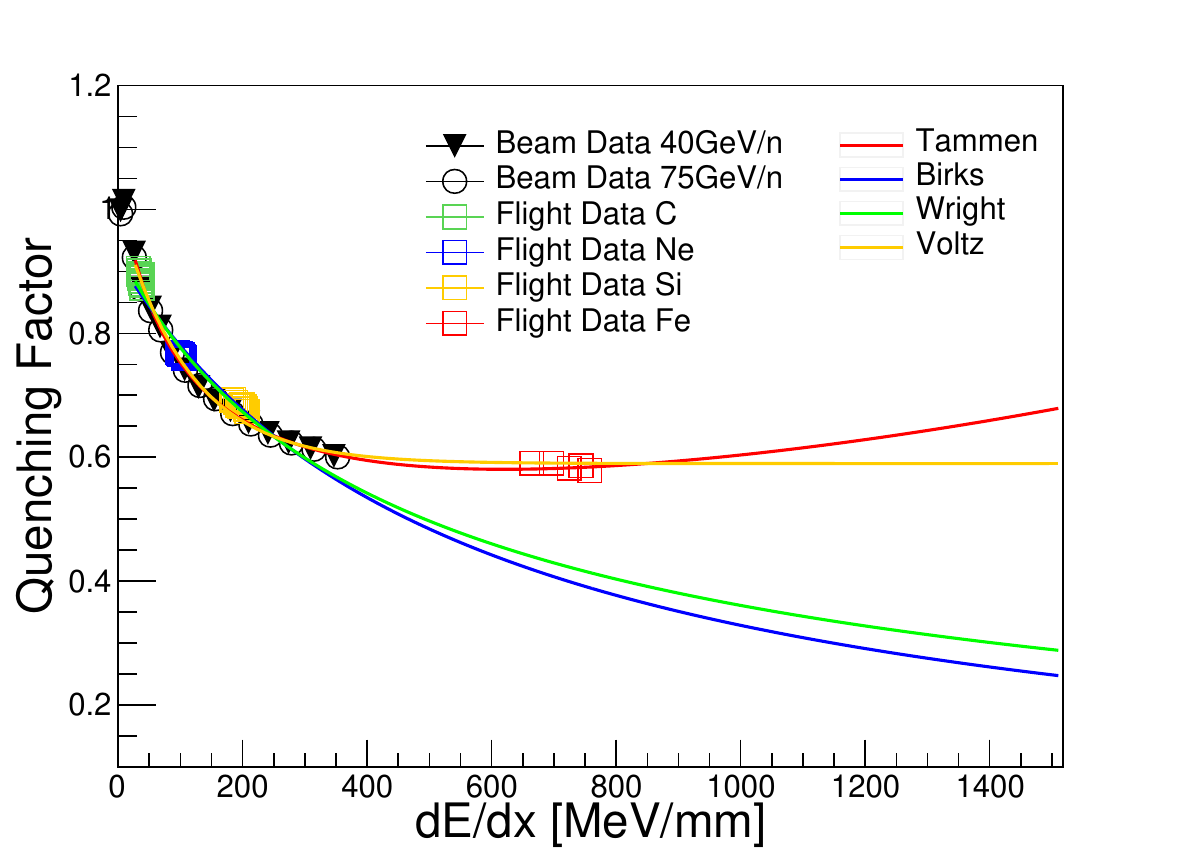}
\caption{The quenching factor as a function of ionization energy density is presented for various nuclei, including cosmic ray carbon, neon, silicon, iron, and beam test helium to argon. The data points used in this study were obtained from Ref.~\cite{2020ITNS...67..939W}. The figure displays four solid lines that correspond to the results of four different quenching function models: Tammen's model (red), Birks's model (blue), Wright's model (green), and Voltz's model (yellow). 
}
\label{quenching}
\end{figure*}

The light yield in the inorganic scintillator is expected to be  proportional to the energy loss per unit distance. However, when the energy loss is relatively large, this relationship becomes impractical and requires modifications. The light yield per unit length in the inorganic scintillator can be calculated as follows:

\begin{equation}
 \frac{dL}{dx}=\epsilon(T)\cdot S\cdot Q(\epsilon(T)),
\label{dLdx}
\end{equation}
where $T$ is the incident particle's kinetic energy, $\epsilon(T)\equiv dE/dx$ is the energy loss, and $Q$ is the dimensionless quenching function.  The quantity S represents the linear relationship between the amount of energy deposited and the resulting light emitted, with a unit of photons per energy.

In this study, we discuss several quenching models, as illustrated in Fig.~\ref{quenching}. The Birks' Law model assumes that the quenching effect is solely due to the recombination of free electrons and ions within the scintillator \cite{birks1951scintillations}. The Birks' Law equation is
\begin{equation}
   Q_{\mathrm{Birks}}(\epsilon)=A/(1+kB\epsilon),
  \label{Birks}
\end{equation}
where $A$ and $kB$ are empirically determined constants that depend on the scintillator material. The kB is called Birks' coefficient, and its unit is distance per energy.

Tammen et al. (2015) enhanced Birks' model by transforming the $kB\epsilon$ term into an exponential function, resulting in a better fit for the model. Its equation is represented as \cite{2015NIMPB.360..129T}
\begin{equation}
Q_{\mathrm{Tammen}}(\epsilon)=\dfrac{A}{1+kB\cdot e^{-\beta\sqrt{kB\epsilon}}\epsilon},
\label{Tammen}
\end{equation}
where $\beta$ is the introduced dimensionless parameter.

Wright et al. \cite{wright1953scintillation} defined the Phenomenological-Logical quenching function, which accounts for both the recombination and diffusion of free electrons and ions within the scintillator . Wright's model can be expressed as:
\begin{equation}
Q_{\mathrm{Wright}} = \dfrac{A}{kB\epsilon}\log(1+kB\epsilon),
 \label{Ph-L}
\end{equation}

Voltz's model, which considers the impact of secondary products on the ionization process and accounts for the branching ratios for different reaction channels, is described in \cite{voltz1966influence}. The mathematical formulation of this model is expressed as follows:
\begin{equation}
Q_{\mathrm{Voltz}} = f+(1-f)e^{-kB(1-f)\epsilon},
\label{Voltz}
\end{equation}
where $f$ denotes the percentage of unquenching deposited energy. It is worth mentioning that alternative models may consider all ionization as resulting from primary reactants, but Voltz's model provides a more comprehensive approach.

Ref.~\cite{2020ITNS...67..939W} investigated the ionization energy response of BGO to incident particles with atomic numbers ranging from one to twenty-six. The study utilized energy loss obtained from beam data and flight data to compare with energy loss generated by GEANT4 simulations of incident particle production. Quenching parameters were calculated from these comparisons. The results show that the quenching effect is positively correlated with the density of ionization energy loss. In our study, we employ four different models to fit the quenching factor and find that the conventional Birks's and Wright's models are insufficient to account for the quenching factors at large ionization densities and Tammen's model appears an unphysical upward curvature. In contrast, Voltz's model provides an excellent fit for the quenching factors in the whole dE/dx range. Therefore, based on Eq.~\ref{dLdx}, the Voltz's model is employed in simulation procedure to tune the ionization energy deposit of each G4-step for all the charged particle species in the BGO calorimeter.

\section{Results}

To validate the performance of the correction on the energy response by including the BGO quenching effect in the simulation procedure, the simulation results of various nuclei based on GEANT4 and FLUKA are compared with the CERN-SPS beam test data of DAMPE, as presented in section \ref{sectionA}. Furthermore, in section \ref{sectionB}, we evaluate the BGO quenching effect on the energy spectral measurement.


\begin{figure*}[htbp]
\centering         
\includegraphics[width=0.32\linewidth]{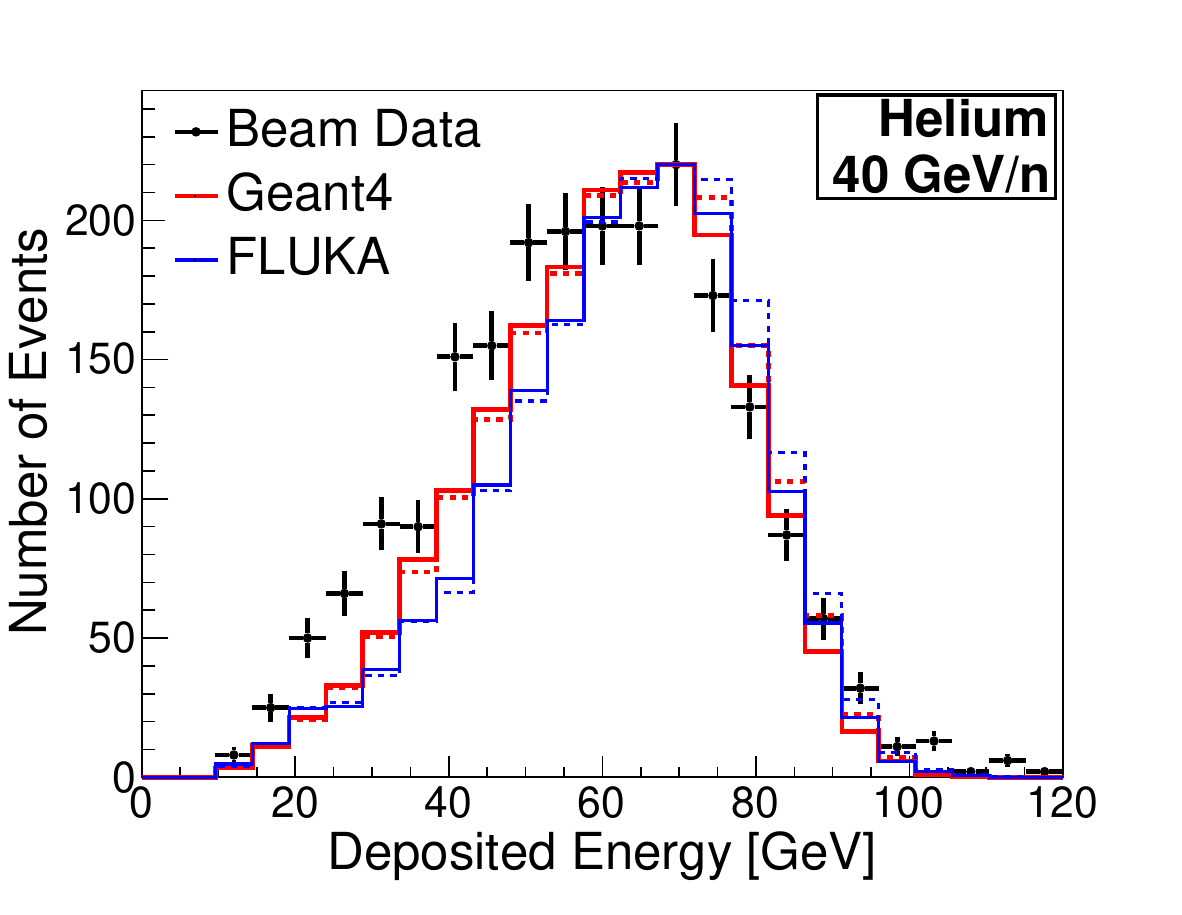}  
\includegraphics[width=0.32\linewidth]{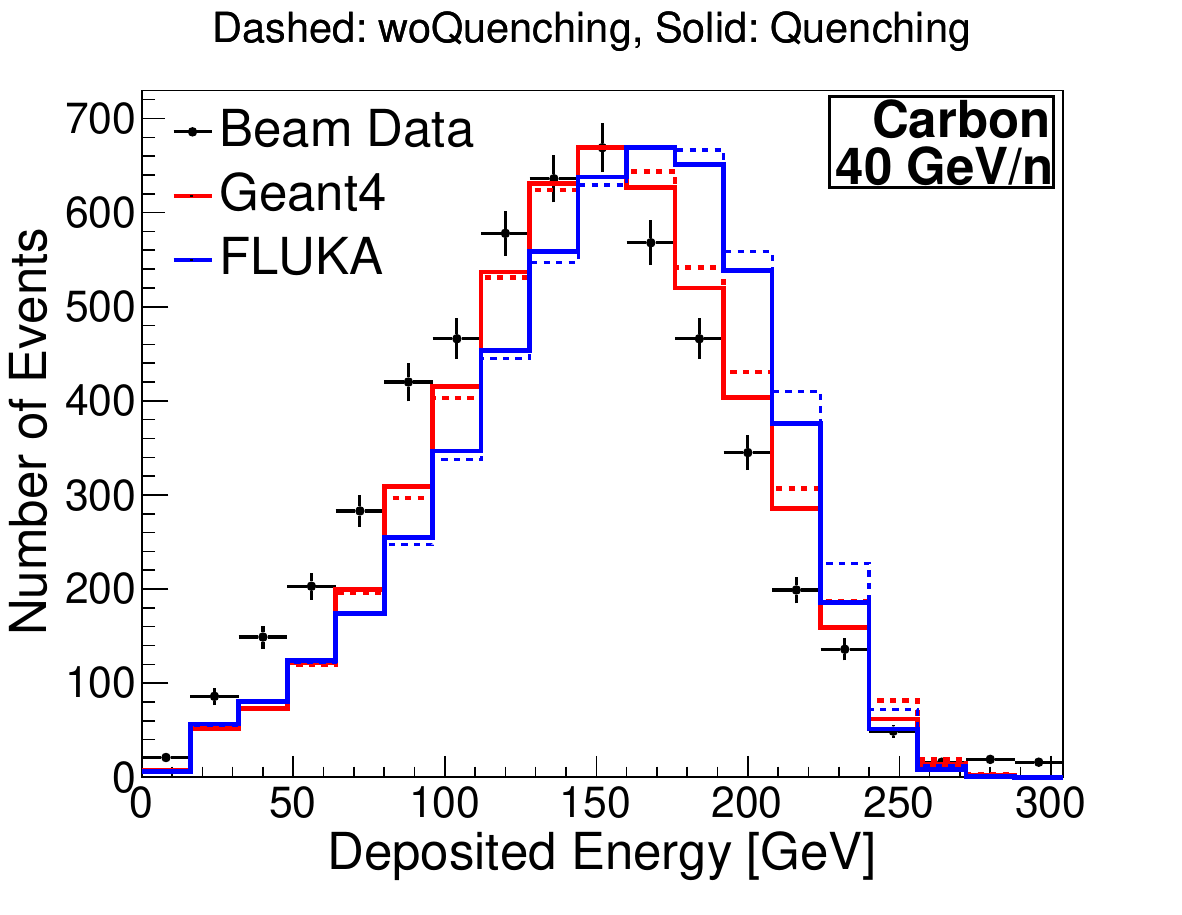}
\includegraphics[width=0.32\linewidth]{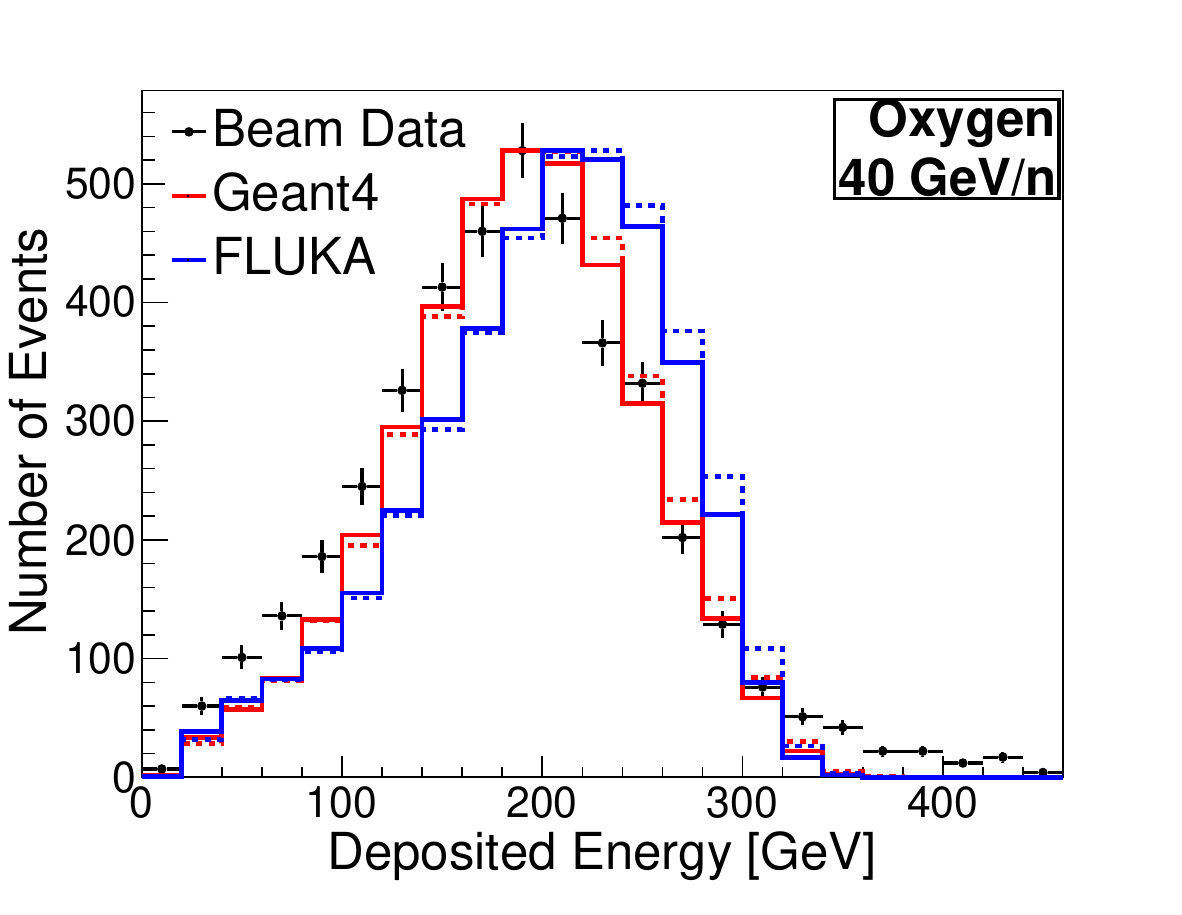}
\includegraphics[width=0.32\linewidth]{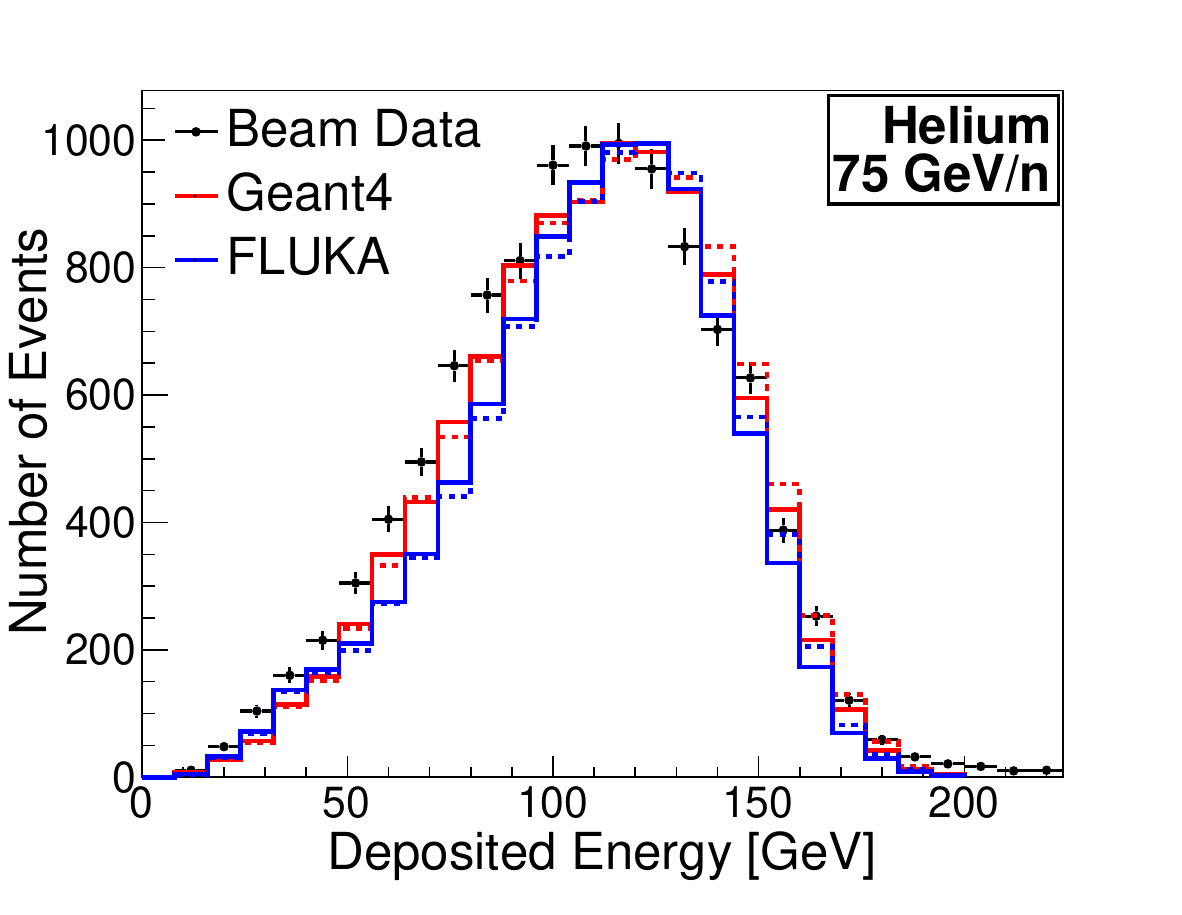} 
\includegraphics[width=0.32\linewidth]{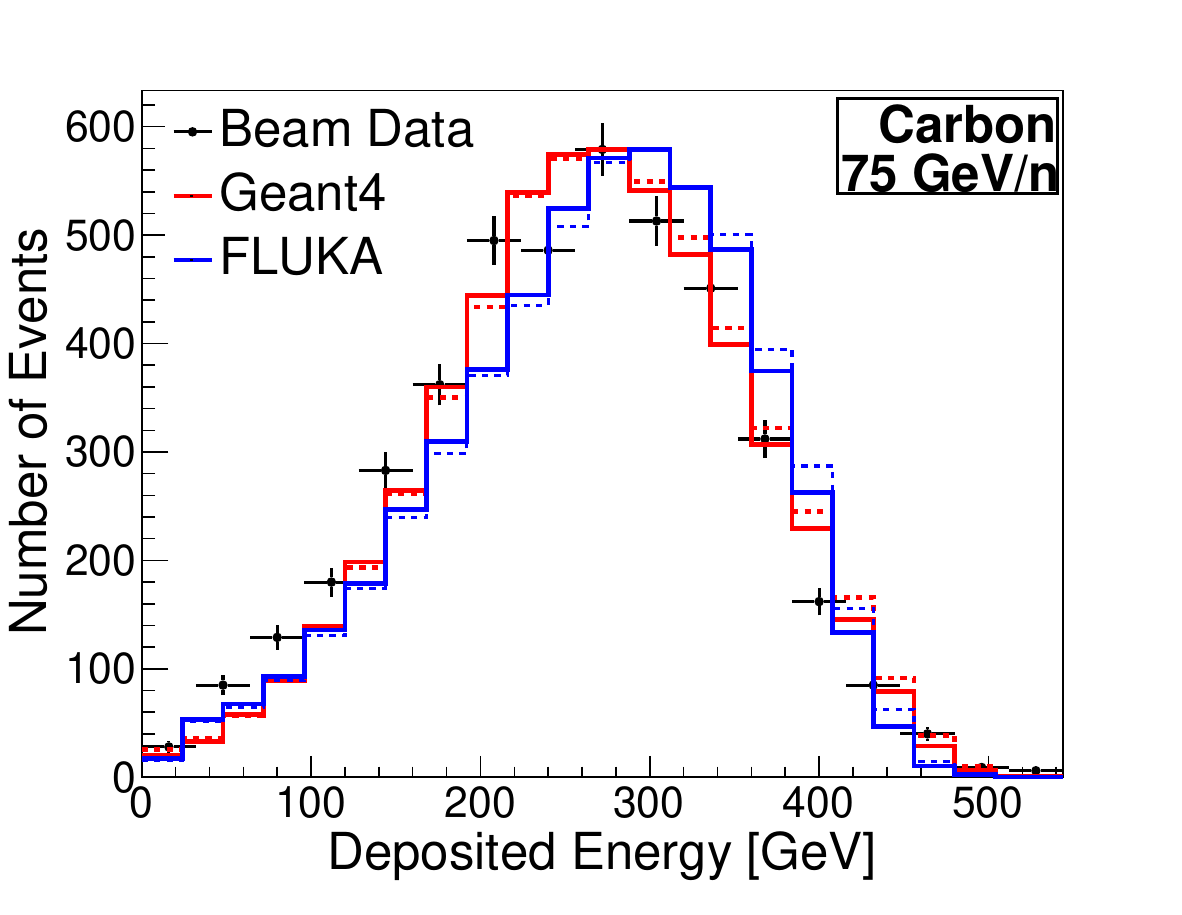} 
\includegraphics[width=0.32\linewidth]{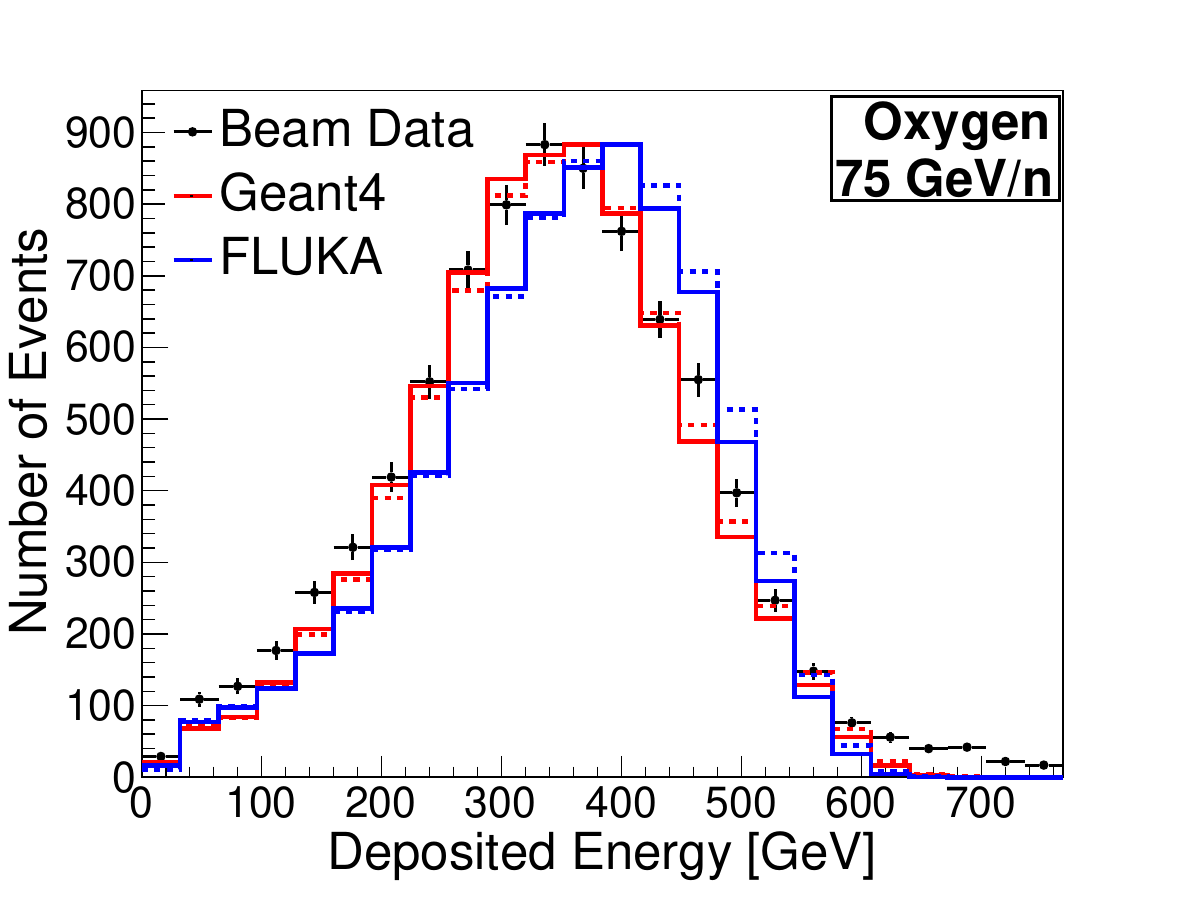} 
\caption{The comparisons of the the energy response for helium, carbon, and oxygen particles between the beam test data (black dot) and the simulations with GEANT4 (blue line) and FLUKA (red line). The deposited energy distributions with and without accounting for the quenching effect are shown respectively in solid and dashed lines.}
\label{BGOQuenching}
\end{figure*}

\begin{figure*}[htbp]
\centering
\includegraphics[width=.49\textwidth]{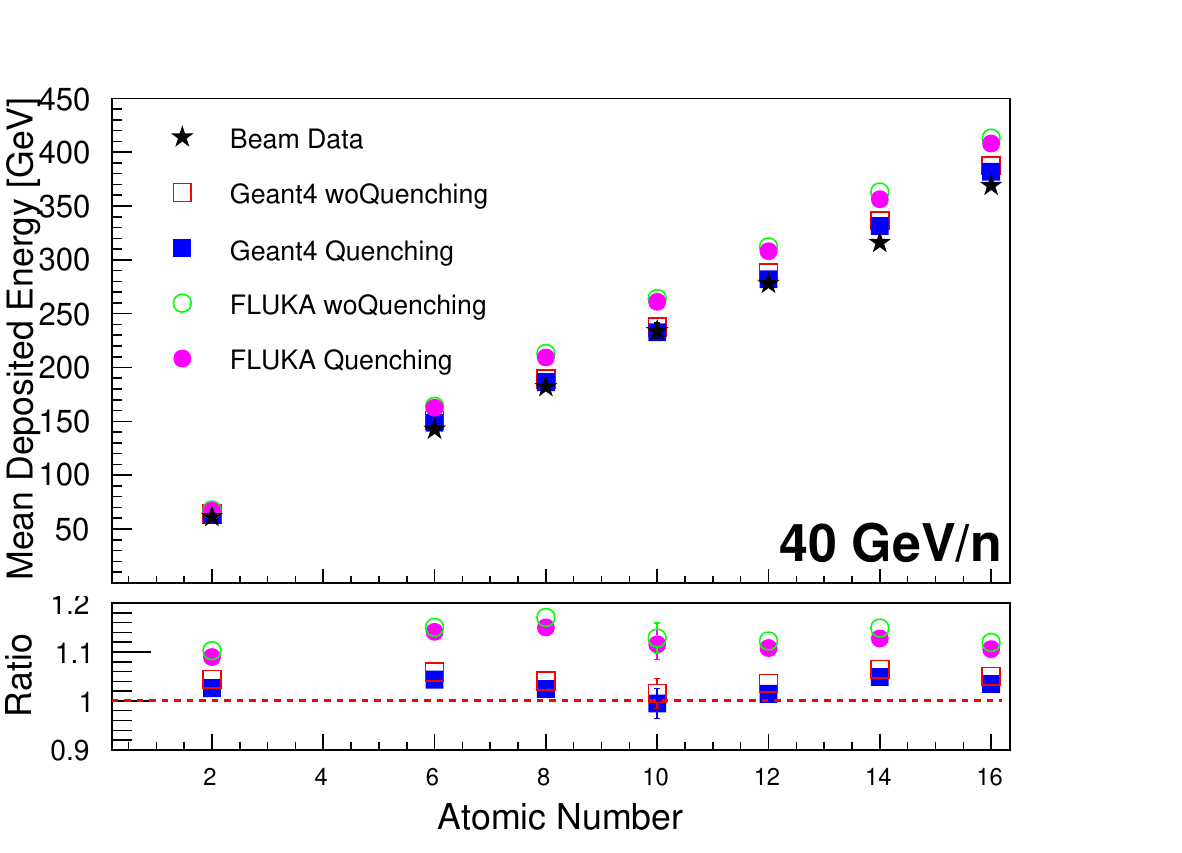}
\includegraphics[width=.49\textwidth]{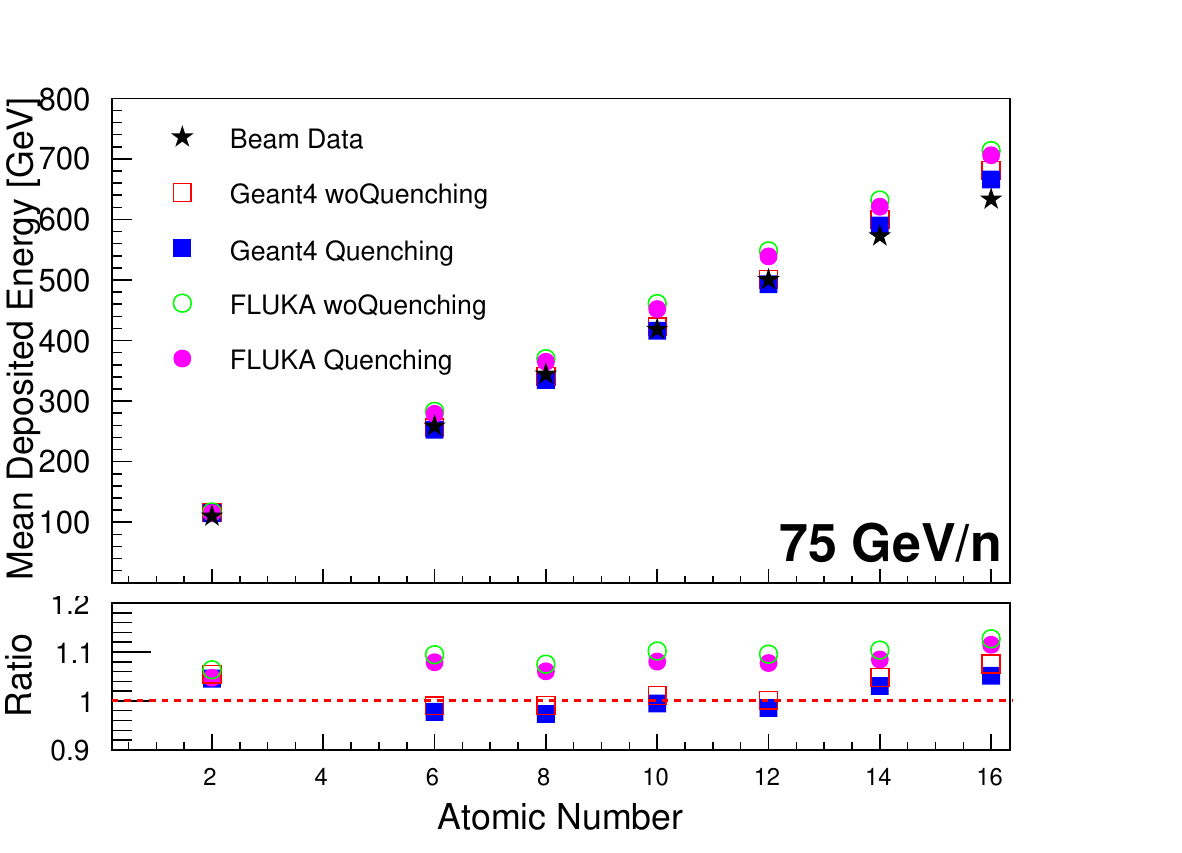}
\caption{The mean deposited energy as a function of the atomic number for kinetic energies of 40 GeV/n (left) and 75 GeV/n (right). The beam test data are shown in black stars. The simulated results without the quenching correction are shown in red squares (GEANT4) and green circles (FLUKA), while the ones with the quenching correction are shown in solid blue squares (GEANT4) and magenta dots (FLUKA), respectively. The ratios of the simulations to the beam test data are shown correspondingly in the bottom panels.}
\label{BgoMeans}
\end{figure*}

\subsection{BGO quenching effect on energy response}
\label{sectionA}

Due the limited thickness of the BGO calorimeter (about 1.6 nuclear interaction lengths) and the presence of muon and neutrino components in hadronic showers, the heavy energy leakage results ein a limited energy resolution for nuclei measurements. An unfolding procedure is thus necessary to account for the bin-to-bin migration effect based on the energy response matrix obtained from the simulations. However, various hadronic interaction models provide quite different energy response matrices, which induces a significant systematic uncertainty on the spectral measurement \cite{2019SciA....5.3793A}.  Before launch, an extensive test of the DAMPE detector was conducted at the H8 beam line of the CERN Super-protons-Synchrotron (SPS) during 2014-2015. Argon-40 particles with primary energy of 40 and 75 GeV/n were targeted at a 40 mm polyethylene target, producing secondary particles with an atomic mass to atomic number ratio of 2.
The energies of these secondary particles are thereby approximately equal to 40 GeV/n and 75 GeV/n. 
The simulations based on GEANT4 and FLUKA are conducted separately with the identical conditions to the beam test, including the incident energy, hit point, direction, and event selections.

In Fig.~\ref{BGOQuenching}, we compare the energy deposition  distributions of the BGO calorimeter in the beam test with the simulated results using GEANT4 and FLUKA, for helium, carbon, and oxygen with kinectic energies of 40 GeV/n and 75 GeV/n. The red and blue lines represent the GEANT4 and FLUKA simulations, and the solid and dashed lines represent the energy deposition with and without considering the quenching effect, respectively. The black dots correspond to the beam test data. The comparisons indicate that both of the shape and the peak of the energy response from GEANT4 are well consistent with the beam test data for all the three ions. While the energy responses from FLUKA are systematically larger than the beam test results. Such a discrepancy between GEANT4 and FLUKA could be attributed to their different branching ratio of the $\pi^0$ secondary in the hadronic interactions between the incident particles and BGO.

In our analysis, we not only compared the shape and peak of the energy response from the simulated and beam tests but also assessed the ratio of the average deposition energy between them. The study involved particles of helium, carbon, nitrogen, oxygen, neon, magnesium, silicon, and sulfur, each with kinetic energies of 40 GeV/n and 75 GeV/n. In Fig.~\ref{BgoMeans}, the upper figure shows that the mean deposition energy of GEANT4, when accounting for the quenching effect, closely matches the beam test result. The lower figure shows the ratios of the mean deposited energy obtained from GEANT4 and FLUKA simulations to that of the beam test, while also accounting for the quenching effect. The results indicate that the GEANT4 simulations show a much better agreement with the beam test outcomes compared with the FLUKA ones.In particular, the BGO quenching correction results in a decrease of  approximately 2.0$\%$ (1.8$\%$) on the mean deposited energy value for 40 GeV/n (75 GeV/n) particles. By accounting for the BGO quenching effect, the simulation would achieve a better performance of the energy response, which is very important for the precise spectral measurement.

\subsection{BGO quenching effect on spectral measurements}
\label{sectionB}

\begin{figure*}[htbp]
\centering     
\includegraphics[width=0.32\linewidth]{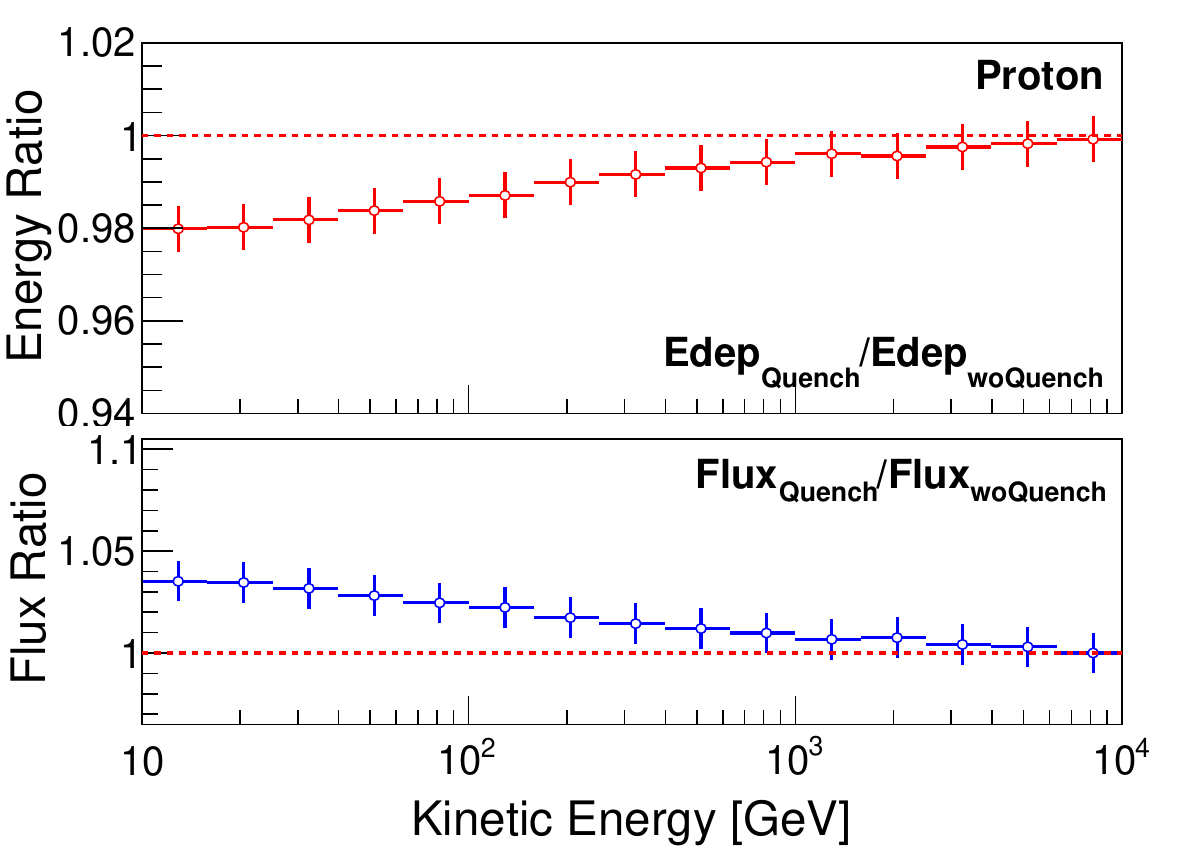} 
\includegraphics[width=0.32\linewidth]{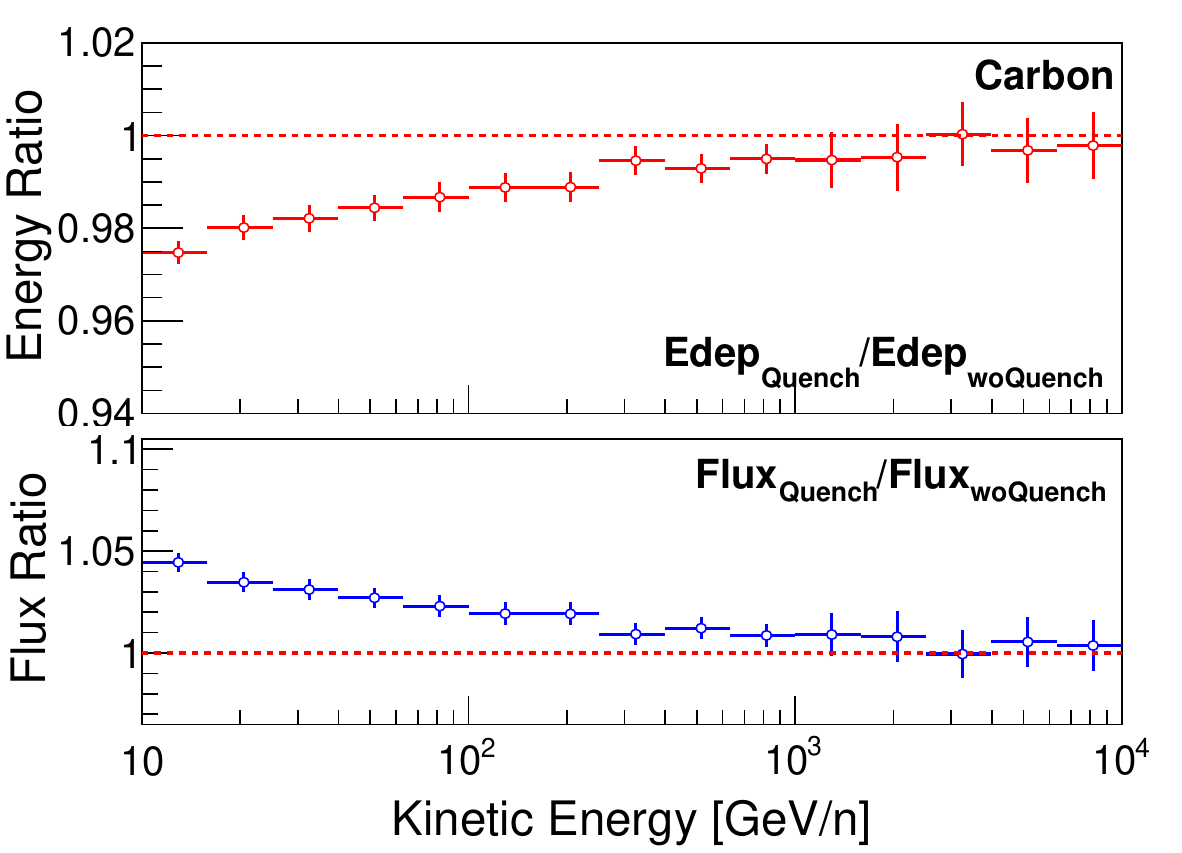}          
\includegraphics[width=0.32\linewidth]{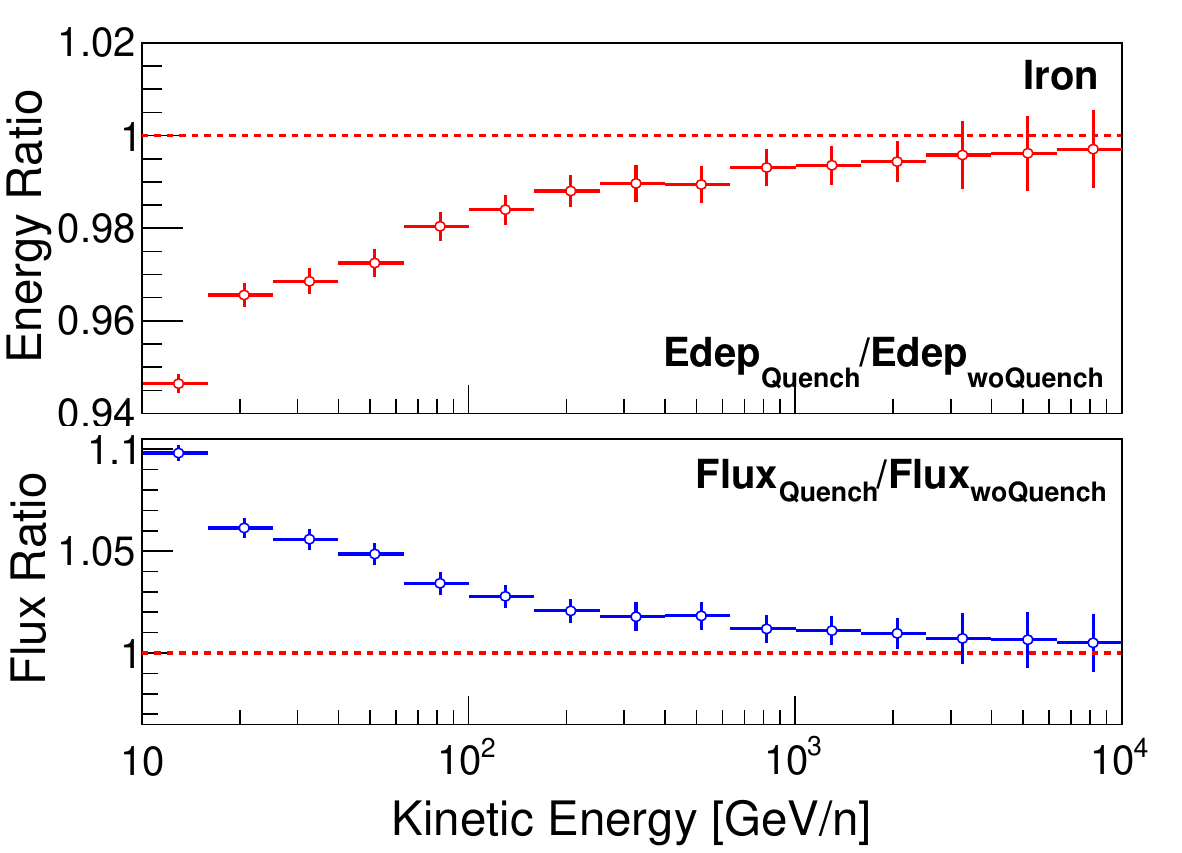}  
\caption{The ratio of the energy deposition (top) and the calculated flux (bottom) with and without the BGO quenching effect as a function of the kenetic energy. The results for proton, carbon, and iron are shown in leftmost, middle and rightmost plots, respectively. 
}
\label{EdepMeans}
\end{figure*}

The flux of cosmic rays, represented by $F(E)$, describes the intensity of cosmic rays by indicating the number of cosmic rays that pass through a unit area per unit of time and unit solid angle. The energy distribution of cosmic rays follows a power-law spectrum, with $F(E)$ being proportional to $E^{-\gamma}$. The spectral index $\gamma$ is a vital parameter for characterizing cosmic rays' energy distribution. The flux equation is expressed as follows:

\begin{equation}
F(E_i,E_i+\Delta E_i) = \frac{N_{{\rm inc},i}}{A_{{\rm eff},i}~\Delta E_i
~T_{\rm exp}};~~~
 N_{{\rm inc},i}=\sum_{j=1}^{n}\mathbf{M}_{ij}N_{{\rm dep},j}\,,
\label{eq-flux}
\end{equation}
where $A_{{\rm eff}, i}$ represents the effective acceptance and $T_{\rm exp}$ represents the exposure time. $\Delta E_i$ is the width of the energy bin, $N_{{\rm inc},i}$ is the number of events in the $i$th incident energy bin, and $N_{{\rm dep},j}$ is the number of events in the $j$th deposited energy bin. To estimate the BGO quenching effect on the spectral measurement, a pseudo deposited energy distribution $N^{*}_{{\rm dep},j}$ is firstly generated to illustrate the experimental measurement by assuming an $E^{2.7}$ incident spectrum in the simulation with the BGO quenching effect. Such a pseudo $N^{*}_{{\rm dep},j}$ is then applied to reconstruct the incident energy spectrum based on the response matrices ($\mathbf{M}{ij}$) with and without the BGO quenching effect, via the Bayesian unfolding approach\cite{d1995multidimensional}. By comparing the output incident spectra, the performance of the BGO quenching correction on the flux measurement is obtained.

In this work, we analyzed the effect of the BGO quenching correction on energy measurement and flux measurement based on the GEANT4 simulations. The performances for proton (left), carbon (middle) and iron (right) are shown in Fig.\ref{EdepMeans}. The top panels of Fig.\ref{EdepMeans} present the ratios of the mean energy deposition with and without the BGO quenching correction. The BGO quenching correction results in a decrease of the measured energy deposition by approximately 2.0$\%$, 2.5$\%$ and 5.4$\%$ at 10 GeV/n for proton, carbon and iron, respectively. For energies above 1 TeV/n, the ratio drops down to within 1$\%$, indicating that the BGO quenching correction is less significant in high energy range. In particular, the effect of iron is much heavier than the ones of proton and carbon, for the same kinetic energy. The bottom panels of Fig.\ref{EdepMeans} show the flux ratios with and without the BGO quenching correction. As expected, the correction of the unfolded flux shows a contrary trend compared with the one of the energy deposition. The correction for proton is approximately 2.0 $\%$ at 10 GeV, which is smaller the value of 4.3$\%$ reported in \cite{2022JInst..17P8014A}. Such a discrepancy would be primarily due to the different BGO quenching models (factors) employed in the two analyses. As for carbon and iron, the BGO quenching correction results in an increase of the flux measurement by approximately 4.4$\%$ and 10.0$\%$ at 10 GeV/n. Also, in high energy range above 1 TeV/n, the correction on the flux measurement becomes less important. Additionally, the BGO quenching correction on the energy measurement of electron is found to be smaller than 0.1$\%$ for energies above 10 GeV, which is negligible for the spectral measurement.

\section{Conclusion}

The accurate spectral measurement of cosmic-ray nuclei with DAMPE, a calorimeter-based experiment, requires a precise instrument response which can only be obtained from the Monte Carlo simulation of the interaction between the incident particle and and the detector material. However, the non-linear fluorescence response of BGO, caused by the quenching effect, results in an energy discrepancy between the simulation and the experimental condition. To address this issue, we test four different quenching models and select the most appropriate model to correct the energy response in the simulation procedure. It shows that GEANT4 provides a consistent energy response for most nuclei with the beam test data, while FLUKA's energy response is generally larger. Also, we estimate the impact of the BGO quenching effect on the energy spectral measurement of various nuclei. The results demonstrate that the quenching effect in BGO reduces the energy deposition of protons, carbon, and iron by approximately 2$\%$, 2.5$\%$, and 5.4$\%$, respectively, at an incident energy of 10 GeV/n. Moreover, the reduction remains below 1$\%$ for energies above 1 TeV/n. Correspondingly, the unfolding spectra show significant increases of about 3.5$\%$, 4.4$\%$, and 10.0$\%$ at 10 GeV/n for protons, carbon, and iron.  This work suggests that the quenching effect of the inorganic scintillator, such as BGO, need to be carefully estimated to obtain a precise energy response of the calorimeter.

\section{Acknowledgements}
This work is supported by the National Key Research and Development Program of China 
(No. 2022YFF0503302), the National Natural Science Foundation of China (No. 12220101003, 
No. 12003076, No. 12275266), the Chinese Academy of Sciences (CAS) Project for Young Scientists in 
Basic Research (No. YSBR-061), the Youth Innovation Promotion Association of CAS (No. 2022320), 
and the Natural Science Foundation of Jiangsu Province (No. BK20201107).

\bibliographystyle{elsarticle-num}
\bibliography{references}
\end{document}